 \documentclass[manuscript]{emulateapj}

\usepackage{apjfonts}

\slugcomment{to appear in the Astrophysical Journal}

\shorttitle{Light Curve Model of V445 Pup }
\shortauthors{Kato et al.}

\begin{document}

\title{HELIUM NOVA ON A VERY MASSIVE WHITE DWARF -- A LIGHT 
CURVE MODEL OF V445 PUPPIS (2000) REVISED }


\author{Mariko Kato}
\affil{Department of Astronomy, Keio University,
Hiyoshi, Kouhoku-ku, Yokohama 223-8521, Japan}
\email{mariko@educ.cc.keio.ac.jp}

\author{Izumi Hachisu}
\affil{Department of Earth Science and Astronomy,
College of Arts and Sciences, University of Tokyo,
Komaba, Meguro-ku, Tokyo 153-8902, Japan}
\email{hachisu@ea.c.u-tokyo.ac.jp}

\author{Seiichiro Kiyota}
\affil{4-405-1003 Matsushiro, Tsukuba 305-0035, Japan}
\email{skiyota@nias.affrc.go.jp}

\and 

\author{Hideyuki Saio}
\affil{Astronomical Institute, Graduate School of Science, Tohoku University, 
Sendai 980-8578, Japan}
\email{saio@astr.tohoku.ac.jp}




\begin{abstract}
V445 Pup (2000) is a unique object identified as a helium nova.
Color indexes during the outburst are consistent with those of
free-free emission.  We present a free-free emission dominated
light curve model of V445 Pup on the basis of the optically thick wind theory. 
Our light curve fitting shows that  (1) the white dwarf (WD) mass
is very massive ($M_{\rm WD} \gtrsim 1.35 ~M_\sun$), and
(2) a half of the accreted matter remains on the WD, both of which 
suggest that the increasing WD mass. Therefore, V445 Pup is a strong candidate of
Type Ia supernova progenitor.  The estimated distance to V445 Pup is
now consistent with the recent observational suggestions,
$3.5 \lesssim d \lesssim 6.5$ kpc.  A helium star companion is consistent with the
brightness of $m_{\rm v}=14.5$ mag just before the outburst, if it is 
a little bit evolved hot ($\log T ~(K) \gtrsim 4.5$) star with
the mass of $M_{\rm He} \gtrsim 0.8 ~M_\sun$.  We then emphasize
importance of observations in the near future quiescent phase
after the thick circumstellar dust dissipates away, especially
its color and magnitude to specify the nature of the companion star.
We have also calculated helium ignition masses for helium shell flashes
against various helium accretion rates and discussed the recurrence period
of helium novae.
\end{abstract}





\section{Introduction}

The outburst of V445 Pup was discovered on 30 December 2000 by Kanatsu 
\citep{kan00}.  The outburst shows unique properties such as 
absence of hydrogen, unusually strong carbon emission lines as well as
strong emission lines of Na, Fe, Ti, Cr, Si, Mg etc. \citep{ash03,iij08}. 
The spectral features resemble those of classical slow novae except
absence of hydrogen and strong emission lines of carbon \citep{iij08}. 
The development of the light curve is very slow (3.3 mag decline
in 7.7 months) with a small outburst amplitude of 6 mag.
From these features, this object has been 
suggested to be the first example of helium novae \citep{ash03,kat03}.

In our earlier work \citep{kat03} we presented a theoretical light curve
model with the assumption of blackbody emission from the photosphere,
which resulted in the best fit model of a very massive white dwarf (WD)
($M_{\rm WD} \gtrsim 1.3 ~M_{\odot}$) and a relatively short distance
of $d \lesssim 1$ kpc. However, \citet{iij08} recently showed that 
there are absorption lines of Na I D1/2 at the velocities of 16.0 and
73.5 km~s$^{-1}$, suggesting that V445 Pup is located in or beyond
the Orion arm and its distance is as large as $3.5 -  6.5$ kpc.
\citet{wou08} also suggested a large distance of $4.9$ kpc. Moreover,
color indexes during the outburst are consistent with those of
free-free emission.  With these new observational aspects we have revised
the light curve model of V445 Pup.

Helium novae were theoretically predicted by \citet{kat89} as a
nova outburst caused by a helium shell flash on a white dwarf. 
Such helium novae have long been a theoretical object
until the first helium nova V445 Pup was discovered in 2000. 
\citet{kat89} assumed two types of helium accretion: one is 
helium accretion from a helium star companion, and the other is
hydrogen-rich matter accretion with a high accretion rate and,
beneath the steady hydrogen burning shell, ash helium accumulates
on the WD.  The latter case is divided into two kinds of systems
depending on whether hydrogen shell burning is steady (stable) or not.
To summarize, helium accretion onto a WD occurs in three types
of systems: (1) helium accretion from a helium star companion; 
(2) steady hydrogen accretion with a rate high enough to keep
steady hydrogen shell burning such as supersoft X-ray sources, e.g., 
RX J0513.9-6951 \citep[e.g.,][]{hac03kb, mcg06}
and CAL 83 \citep[e.g.,][]{sch06},
(3) hydrogen accretion with a relatively low rate resulting in
a recurrent nova such as RS Oph \citep[e.g.,][]{hac07kl}
and U Sco \citep[e.g.,][]{hkkm00}.
In the present paper, we regard that V445 Pup is Case (1), because 
no hydrogen lines were detected \citep{iij08}. 

In helium novae, mass loss owing to optically thick wind is relatively
weak compared to hydrogen novae, and a large part of the accreted
helium burns into carbon and oxygen and accumulates on the WD
\citep{kat99,kat04}.
Our previous model indicated that V445 Pup contains
a massive WD ($M_{\rm WD} \gtrsim 1.3 ~M_{\odot}$) and the WD mass is
increasing through helium shell flashes. 
Therefore, V445 Pup is
a strong candidate of Type Ia supernova progenitors.
Although there are two known evolutional paths of the single
degenerate scenario toward Type Ia supernova \citep[e.g.,][]{hkn99},
helium accreting WDs such as V445 Pup do not belong to either of them.
Therefore, it may indicate the third path to Type Ia supernovae.  
If various binary parameters of V445 Pup are determined,
they provide us important clues to binary evolutions to Type Ia supernovae.

In the next section we introduce our multi-band photometric
observations, which indicates that free-free emission dominates
in optical and near infrared.  In \S 3 we present
our free-free emission dominated light curve model and its
application to V445 Pup.  Then we discuss brightness
in quiescence in \S 4.
Discussion and conclusions follow in \S 5 and \S 6, respectively.

\section{Observation}

Shortly after the discovery of V445 Pup outburst, one of us (S.K.)
started multi-band photometry with a 25.4 cm telescope 
[focal length $=1600$~mm, CCD camera $=$ Apogee AP-7 (SITE SIA502AB
of $512 \times 512$ pixels)].  $V$ and $R_{\rm c}$ magnitudes are
obtained from January 4 to May 6, 2001, and $I_{\rm c}$ from
January 4, 2001 to January 15, 2007, with the comparison star, 
TYC 6543-2917-1 ($V=8.74$, $B-V=0.304$).  All of our data can be taken
from the data archive of Variable Star Observers League in Japan
(VSOLJ)\footnote{http://vsolj.cetus-net.org/}.


\begin{figure}
\epsscale{1.15}
\plotone{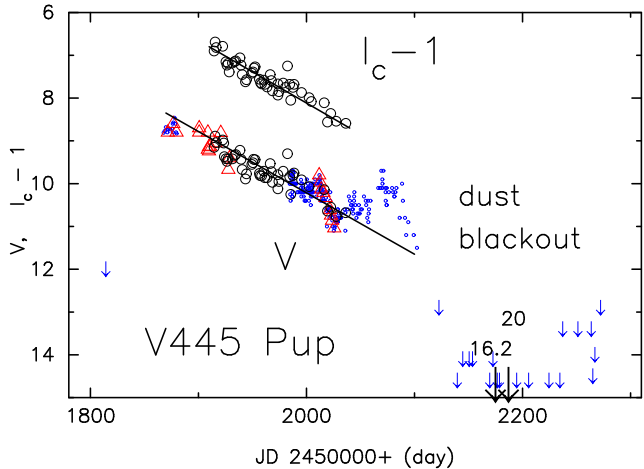}
\caption{
$V$ and $I_{\rm c}$ light curves of V445 Pup (large open circles).
The $I_{\rm c}$ data are shifted up by one magnitude in order to
separate it from the $V$ light curve.  Open triangles denote
$V$ magnitudes (IAUC No. 7552,
7553, 7557, 7559, 7569, 7620).  Small open circles and downward arrows 
are taken from AAVSO to show very early and later phases of the outburst. 
Large downward arrows show an upper limit observation of   
$I_{\rm c} > 16.2$ (this work), $V > 20$ and $I > 19.5$ \citep{hen01}. 
Straight lines indicate the average decline rates of
$1.9 ~{\rm mag} / 130 $~days $= 0.0146 ~{\rm mag}~{\rm day}^{-1}$ and 
$3.3 ~{\rm mag} / 230$ days $= 0.0143 ~{\rm mag}~{\rm day}^{-1}$ 
for  $I_{\rm c}$ and $V$, respectively. 
}
\label{lightobs.only}
\end{figure}

Figure \ref{lightobs.only} shows our $V$ and $I_{\rm c}$ magnitudes
as well as other observations taken from literature. 
These two light curves show very slow evolution
($\sim 0.014$ mag day$^{-1}$) followed by an oscillatory behavior
in the $V$ magnitude before it quickly darkened by dust blackout
on about JD 2452100, i.e., 7.5 months after the discovery
\citep{hen01, wou02, ash03}. 
Here, we assume the decline rate of the light curves as shown
in Figure \ref{lightobs.only} ignoring the later oscillatory phase
(JD 2452040 -- 2452100) because we assume steady-state in the
nova envelope as mentioned below and, as a result, our theoretical
light curve cannot treat unsteady oscillations.


\begin{figure}
\epsscale{1.15}
\plotone{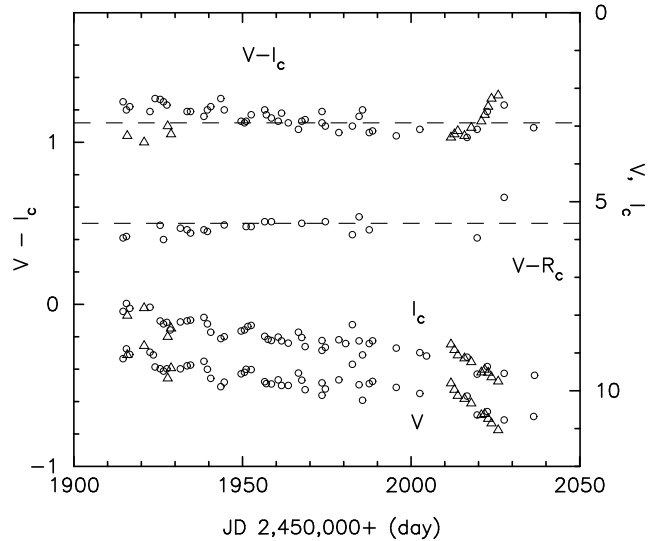}
\caption{
Color indexes of $V-I_{\rm c}$ and $V-R_{\rm c}$ as well as $I_{\rm c}$
and $V$.  Open circles: this work. Open triangles: Gilmore
(IAUC 7559, 7569) and Gilmore \& Kilmartin (IAUC 7620).
The horizontal dashed lines indicate the mean values of
$V-R_{\rm c}=0.5$ and $V-I_{\rm c}=1.12$.
}
\label{VmIc}
\end{figure}

Figure \ref{VmIc} shows evolution of color indexes $V-I_{\rm c}$ and
$V-R_{\rm c}$ as well as $V$ and $I_{\rm c}$ themselves.
We can see that both of $V-I_{\rm c}$ and  $V-R_{\rm c}$ are
almost constant with time, i.e., $\approx 1.12$ and $\approx 0.5$,
respectively, during our observation.  This means that the three 
light curves of $V$, $I_{\rm c}$, and  $R_{\rm c}$ are almost parallel
and each light curve shape is independent of wavelength.  These
are the characteristic properties of free-free emission dominated
light curves as explained below. 

The flux of optically thin free-free emission is inversely proportional to the 
square of wavelength, i.e., $F_\lambda \propto \lambda^{-2}$.
This spectrum shape is unchanged with time although the electron
temperature rises during nova outburst because the emission
coefficient depends only very slightly on the electron temperature
\citep[e.g.,][]{bro70}. 
Therefore, the color indexes of free-free emission dominated
light curves are unchanged with time.  On the other hand,
the color indexes change with time if blackbody emission dominates
because the photospheric temperature rises with time in nova outbursts.
Figures  \ref{lightobs.only} and \ref{VmIc} indicate 
that the emission from V445 Pup is dominated by free-free emission
rather than by blackbody emission.


\begin{figure}
\epsscale{1.15}
\plotone{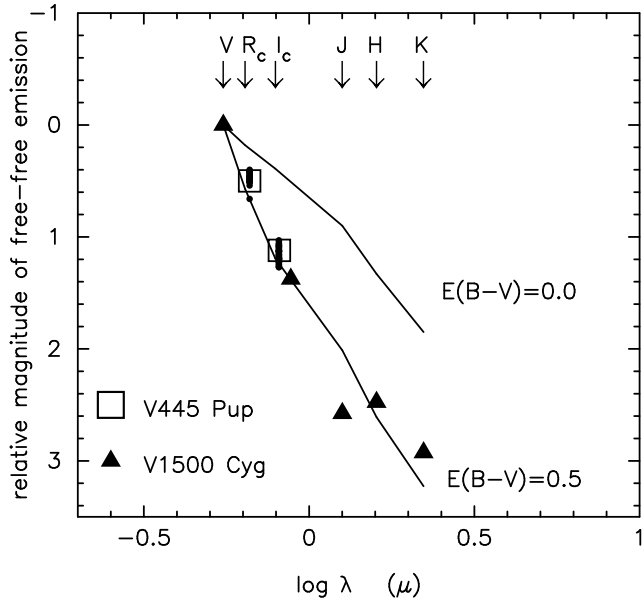}
\caption{
Color indexes against wavelength.  Our observational color indexes
of V445 Pup, $V-R_{\rm c}$ and $V-I_{\rm c}$, are plotted
for individual data (dots) and their mean values of 0.5 and 1.12
(open squares).  The color indexes of V1500 Cyg are added
for $y$ magnitude instead of $V$ (filled triangles)
\citep[taken from][]{hac07kc}.  Arrows indicate the effective
wavelength of six bands of $V$, $R$, $I$, $J$, $H$, and $K$. 
Solid curves denote color indexes of free-free
emission for $E(B-V)=0$ and 0.5.  }
\label{magnitude_calibration}
\end{figure}

The color index of free-free emission is calculated from  
$\lambda F_\lambda \propto \lambda^{-1}$.  
When the reddening is known, the reddened color is obtained as 
\begin{equation}
m_V - m_\lambda = (M_V - M_\lambda)_0 + c_\lambda E(B-V),
\label{color_reddening_relation}
\end{equation}
where $(M_V - M_\lambda)_0$ is the intrinsic color and
$c_\lambda$ is the reddening coefficient 
(these values are tabulated in Table 4 in \citet{hac07kc} 
for five colors of $V-R$, $V-I$, $V-J$, $V-H$, and $V-K$).  
Figure \ref{magnitude_calibration} shows the color indexes relative
to $V$ for two cases of reddening, i.e., $E(B-V)=0$ and 0.5.

For comparison, we have added four relative colors of V1500 Cyg 
\citep{hac06} because its continuum flux is known to be that of
free-free emission \citep{gal76} except for the first few days
after the optical peak.  Here we use Str\"omgren $y$ magnitude
instead of $V$, because the intermediate-wide $y$ band
is almost emission-line free.
Three ($y-I$, $y-H$, and $y-K$) of four color indexes are consistent
with those of color indexes with $E(B-V)=0.5$ \citep{fer77, hac06}.
The $J$ band is strongly contaminated with strong emission lines
of \ion{He}{1} \citep[e.g.][]{bla76, kol76, she76}.
This is the reason why the $y-J$ color index deviates from
that of free-free emission \citep*[see][for more details]{hac07kc}.

Our observed color indexes are shown in Figure \ref{magnitude_calibration}. 
The open squares are centered at the mean values of $V-R_{\rm c}=0.5$ and 
$V-I_{\rm c}=1.12$, whereas dots represent individual observations. 
In V445 Pup, no strong emission lines dominate the continuum 
in $V$, $R$, and $I$ bands \citep{kam02, iij08}.  Therefore we conclude
that the color indexes of V445 Pup are consistent with those of
free-free emission with the reddening of $E(B-V)=0.51$ \citep{iij08}.

\section{Modeling of Nova Light Curves}
\label{model_nova_outburst}

\subsection{Basic Model}
After a thermonuclear runaway sets in on the surface of a WD,
the envelope expands to a giant size and the optical luminosity
reaches its maximum.  Optically thick winds occur and
the envelope reaches a steady state.  Using the same method
and numerical techniques as in \citet{kat94h,kat03}, 
we have followed evolution of a nova by connecting steady state
solutions along the sequence of decreasing envelope mass.
We have solved the equations of motion, continuity, radiative diffusion,
and conservation of energy, from the bottom of the helium envelope
through the photosphere.
The winds are accelerated deep inside the photosphere.
Updated OPAL opacities are used \citep{igl96}.  
As one of the boundary conditions for our numeral code, we assume that
photons are emitted at the photosphere as a blackbody with a photospheric
temperature, $T_{\rm ph}$. \citet{kat03} calculated the visual magnitude
$M_V$ on the basis of blackbody emission and constructed the light curves.
In the present work, we calculate free-free emission dominated
light curves.  The envelope structure, wind mass loss rate,
photospheric temperature, and photospheric radius of the WD envelope
are essentially the same as those in the previous model.

The flux of free-free emission from the optically thin region outside
the photosphere dominates in the continuum flux of optical and
near infrared wavelength region and is approximated by
\begin{equation}
F_\nu \propto \int N_e N_i d V
\propto \int_{R_{\rm ph}}^\infty {\dot M_{\rm wind}^2
\over {v_{\rm wind}^2 r^4}} r^2 dr
\propto {\dot M_{\rm wind}^2 \over {v_{\rm ph}^2 R_{\rm ph}}},
\label{free-free-wind}
\end{equation}
where $F_\nu$ is the flux at the frequency $\nu$, $N_e$ and $N_i$ 
are the number densities of electrons and ions, respectively,
$V$ is the volume, $R_{\rm ph}$ is the photospheric radius, 
$\dot M_{\rm wind}$ is the wind mass loss rate, and $v_{\rm ph}$ is
the photospheric velocity \citep{hac06, hac07kc}. 

The proportionality constant in equation (\ref{free-free-wind}) cannot be 
determined a priori because we do not calculate radiative transfer
outside the photosphere: These proportionality constant are determined
by fitting with observational data \citep{hac06, hac07k, hac07kc}.

\subsection{Free-free Light Curve and WD Mass}
We have calculated free-free emission dominated light curves of V445 Pup 
for WD masses of 1.2, 1.3, 1.33, 1.35, 1.37, and $1.377 ~M_{\odot}$.
The last one is the upper limit of the mass-accreting WDs \citep{nom84}.
The adopted WD radius and chemical composition are listed in Table
\ref{composition}.  Here we assume that chemical composition is constant
throughout the envelope. 


\begin{deluxetable}{llll}
\tabletypesize{\scriptsize}
\tablecaption{Model Parameters\tablenotemark{a}
 \label{composition}}
\tablewidth{0pt}
\tablehead{
\colhead{$M_{\rm WD}$} & \colhead{$\log R_{\rm WD}$} & \colhead{$Y$} &
\colhead{$X_{\rm C+O}$} \cr
\colhead{($M_{\odot}$)} & \colhead{($R_{\odot}$)} & \colhead{}
 & \colhead{}
}
\startdata
1.2 & -2.193  & 0.68 & 0.3 \cr
1.3 & -2.33   & 0.48 & 0.5 \cr
1.33 & -2.417 & 0.38 & 0.6 \cr
1.35 & -2.468 & 0.38 & 0.6  \cr
1.37 & -2.535 & 0.58 & 0.4  \cr
1.37 & -2.535 & 0.38 & 0.6  \cr
1.37 & -2.535 & 0.18 & 0.8  \cr
1.377 & -2.56 & 0.38 & 0.6
\enddata
\tablenotetext{a}{The heavy element content is assumed to be $Z=0.02$ 
that includes carbon and oxygen by the solar ratio.
}
\end{deluxetable}

%

\begin{figure}
\epsscale{1.15}
\plotone{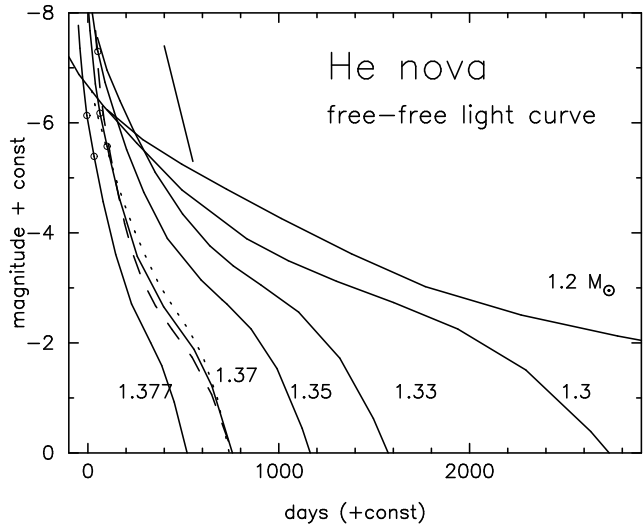}
\caption{
Free-free emission dominated light curves for various WD masses.
The chemical composition is assumed to be $Y=0.38$,
$X_{\rm C+O}= 0.6$, and $Z=0.02$
throughout the envelope.  The WD mass is attached to each curve.
The dashed and dotted curves denote $M_{\rm WD}=1.37  ~M_\sun$ model
but with different chemical composition of $X_{\rm C+O}= 0.4$ and 0.8,
respectively.  Small open circles denote the starting points of
light curve fitting in Fig. \ref{lightcurve_fitting}:
upper and lower points of $M_{\rm WD}= 1.37 ~M_\sun$ (solid 
line) correspond to Models 4 and 3
and those of 1.377 $M_\odot$ correspond to Models 8 and 7, respectively. 
A short solid line indicates the decline rate of observed $V$ data in Fig.
\ref{lightobs.only} (3.3 mag / 230 days).
}
\label{lightcurve}
\end{figure}

Figure \ref{lightcurve} shows the calculated free-free light curves. 
More massive WDs show faster decline.  This is mainly because
the more massive WDs have the less massive envelope and the smaller
mass envelope is quickly taken off by the wind. 
The figure also shows two additional models of $M_{\rm WD}= 1.37 
~M_\sun$ with different compositions of $X_{\rm C+O} =0.4$ and 0.8. 
These models show almost similar decline rates to that of 
$M_{\rm WD}= 1.37 ~M_\sun$ with $X_{\rm C+O}=0.6$.

The starting point of our model light curve depends on the initial
envelope mass.  When an initial envelope mass is given,
our nova model is located somewhere on the light curve.
For a more massive ignition mass, it starts from an upper point of
the light curve.  Then the nova moves rightward with the decreasing 
envelope mass due to wind mass loss and nuclear burning. 

The development of helium shell flashes is much slower than that of
hydrogen shell flashes mainly because of much larger envelope masses
of helium shell flashes \citep{kat94h}. 
The short straight solid line in Figure \ref{lightcurve} represents
the decline rate of $V$ light curve in Figure \ref{lightobs.only}. 
The length of the line indicates the observed period which is relatively
short because of dust blackout 230 days after the optical maximum. 	
We cannot compare the entire evolution period of the light curve 
with the observational data so that fitting leaves some ambiguity
in choosing the best fit model.
Even though, we may conclude that WDs with masses of 
$M_{\rm WD} \lesssim 1.33 ~M_\sun$ are very unlikely
because their light curves are too slow to be comparable with 
the observation.


\begin{deluxetable*}{lllllllllr}
\tabletypesize{\scriptsize}
\tablecaption{Parameters of Fitted Model
 \label{fitting_parameter}}
\tablewidth{0pt}
\tablehead{
\colhead{model } & \colhead{$M_{\rm WD}$} & \colhead{$Y$}
& \colhead{$X_{\rm C+O}$} &
\colhead{$\Delta M_{\rm He,ig}$\tablenotemark{a}}
& \colhead{$\Delta M_{\rm ej}$\tablenotemark{b}}
& \colhead{$\eta_{\rm He}$\tablenotemark{c}}& \colhead{Distance}
& \colhead{$\dot M_{\rm He}$\tablenotemark{d}} & \colhead{$\tau_{\rm rec}$\tablenotemark{d}} \cr
\colhead{No. } & \colhead{($M_\sun$)} & \colhead{} & \colhead{} 
&\colhead{($M_{\odot}$)} &\colhead{($M_{\odot}$)} & \colhead{}
& \colhead{(kpc)} & \colhead{($M_\sun$~yr$^{-1}$)}
& \colhead{(yr)}
}
\startdata
1& 1.35 & 0.38  & 0.6 &3.3E-4 &1.8E-4 & 0.45 & 6.6 & 6.E-8 & 5000 \cr
2& 1.37 & 0.58  & 0.4 & 1.6E-4 &9.2E-5 & 0.42  &5.0  &7.E-8  &2000 \cr
3& 1.37 & 0.38  & 0.6 &1.9E-4 &8.7E-5 & 0.53 & 4.8 & 6.E-8 & 3000 \cr
4& 1.37 & 0.38  & 0.6 &2.1E-4 &1.1E-4 & 0.49 & 5.5 & 5.E-8 & 4000 \cr
5& 1.37 & 0.18  & 0.8 &3.5E-4 &1.2E-4 &0.64 & 6.0 & 4.E-8  & 9000\cr
6& 1.37 & 0.18  & 0.8 &3.8E-4 &1.5E-4 &0.61 & 6.9 &4.E-8 &10000 \cr
7& 1.377 & 0.38 & 0.6 &1.8E-4 &7.4E-5 & 0.60 & 4.8 & 5.E-8 & 4000\cr
8& 1.377 & 0.38 & 0.6 &2.2E-4 &9.6E-5 & 0.56 & 5.2 & 4.E-8 & 5000 \cr
\enddata
\tablenotetext{a}{Envelope mass at ignition is assumed to be equal to the 
mass at the optical peak, which we assume as JD 2451872
(28 days before the discovery).}
\tablenotetext{b}{Total mass ejected by the wind}
\tablenotetext{c}{Mass accumulation efficiency}
\tablenotetext{d}{estimated from Fig. \ref{dMenvMacc}}
\end{deluxetable*}

\begin{figure}
\epsscale{1.15}
\plotone{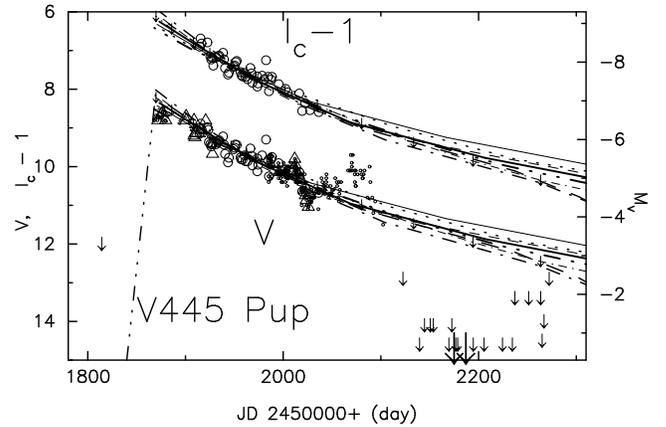}
\caption{
Our model light curves are fitted with the observation. 
Dotted line: Model 1. Thin dashed line: Model 2. Thin solid line: Model 3.
Thick solid line: Model 4.  Dash-three dotted line: Model 5. 
Thin dash-dotted line: Model 6. Dashed line: Model 7.
Dash-dotted line: Model 8. 
Our model $I_{\rm c}$ light curves are
identical with those of $V$ but are lifted up by 1.12 mag
(and further shifted up by 1 mag for $I_{\rm c} - 1$).
The ordinate on the right vertical axis represents
the absolute magnitude of Model 4.  
For the other models, the model light curves are shifted down
by 0.4 mag (Model 1), up by 0.2 mag (Model 2) up by 0.3 mag (Models
3 and 7), down by 0.2 mag (Model 5), down by 0.5 mag (Model 6),
up by 0.1 mag (Model 8).
 The observational data are same as those
in Fig. \ref{lightobs.only}.
}
\label{lightcurve_fitting}
\end{figure}

Figure \ref{lightcurve_fitting} represents the light curve fitting
with observational data.  Here, the model light curves of $I_{\rm c}$ 
are identical to those of $V$ but are lifted up by 1.12 mag, 
which is the mean value of $V-I_{\rm c}$ obtained in Figure \ref{VmIc}.  
It should be noted that, in this figure, these light curves are further
lifted up by 1.0 mag because of $I_c - 1$.


These light curves are a part of the model light curves
in Figure \ref{lightcurve}.
In the early phase of the outburst the light curve declines almost
linearly so that fitting is not unique.  We can fit any part of 
our model light curve if its decline rate is the same as 
$\approx 0.014-0.015 ~{\rm mag}~{\rm day}^{-1}$, for example,
either top or middle part of the same light curve of $1.377 ~M_\sun$
WD in Figure \ref{lightcurve}.  In such a case we show two
possible extreme cases, i.e., the earliest starting point and
the latest one by open circles, as shown in Figure \ref{lightcurve}. 
These model parameters are summarized in Table \ref{fitting_parameter},
where we distinguish the model by model number.

Thus we have selected several light curves which show a reasonable 
agreement with the observation. 
However, we cannot choose the best fit model among them because 
the observed period is too short to discriminate a particular
light curve from others because the difference among them
appears only in a late stage. 

For relatively less massive WDs of $M_{\rm WD} \lesssim 1.33 ~M_{\odot}$,
we cannot have reasonable fits with the observation for any part of 
the model light curves.  Among relatively more massive WDs 
of $M_{\rm WD} \gtrsim 1.35 ~M_{\odot}$ in Table \ref{fitting_parameter},
Model 1 shows slightly slower decline.  Therefore, the 
$1.35 ~M_{\odot}$ WD may be the lowest end for the WD mass.
Thus we may conclude that V445 Pup contains a very massive
WD of $M_{\rm WD} \gtrsim 1.35 ~M_{\odot}$.

\subsection{Mass Accumulation Efficiency}

During helium nova outbursts, a part of the helium envelope is blown
off in winds, while the rest accumulates on the WD.  We define the mass
accumulation efficiency, $\eta_{\rm He}$, as the ratio of the envelope
mass that remains on the WD after the helium nova outburst to the helium
envelope mass at ignition, $\Delta M_{\rm He,ig}$ \citep{kat04}.

The mass accumulation efficiency is estimated as follows. 
We have calculated the mass lost by winds during the outburst,
$\Delta M_{\rm ej}$.  Note that $\Delta M_{\rm ej}$ is the calculated
total ejecta mass which is ejected during the wind phase,  
and not the mass ejected during the observing period which may be
shorter than the wind phase.

The ignition mass is approximated by the envelope mass at the optical
peak,  i.e., $\Delta M_{\rm He,ig} \approx \Delta M_{\rm He,peak}$.
Since the exact time of the optical peak is unknown, we assume
that the optical peak is reached on JD 2451872, i.e., the earliest
prediscovery observation in the brightest stage reported to IAU Circular
No. 7553, 28 days earlier than the discovery. 
The envelope mass at the optical peak is estimated from our
wind solution and is listed as $\Delta M_{\rm He,ig}$
in Table \ref{fitting_parameter}.
The resultant accumulation efficiency,
\begin{equation} 
\eta_{\rm He} \equiv {{\Delta M_{\rm He,peak}-\Delta M_{\rm ej}}
 \over {\Delta M_{\rm He,peak}}},
\end{equation} 
is also listed in Table \ref{fitting_parameter}.
The efficiencies are as high as $\sim 50$\%.  
The WD of V445 Pup is already very massive
($ M_{\rm WD} \gtrsim 1.35 ~M_{\odot}$) and its mass has increased
through helium nova outbursts.  Therefore, V445 Pup is a strong
candidate of Type Ia supernova progenitors.

\section{Quiescent Phase}
Before the 2000 outburst, there was a 14.5 mag star
at the position of V445 Pup (taken from the archive of 
VSNET\footnote{http://vsnet.kusastro.kyoto-u.ac.jp/vsnet/}),
but no bright star has been observed since the dust blackout.
We may regard that 14.5 mag is the preoutburst magnitude of the binary.
There are two possible explanations for this quiescent phase luminosity, 
one is the accretion disk luminosity and the other is the luminosity of
bright companion star. In the following subsections, we discuss how 
these possible sources contribute to the quiescent luminosity.

\subsection{Accretion Disk}
In some nova systems, an accretion disk mainly contributes to the brightness
in their quiescent phase.  If the preoutburst luminosity of V445 Pup
comes from an accretion disk, its absolute magnitude is approximated by 
\begin{eqnarray}
M_V {\rm (obs)}&=& -9.48 -{5\over 3}\log\left({M_{\rm WD}\over M_\sun}
{{\dot M_{\rm acc}}\over {M_\sun ~{\rm yr}^{-1}}} \right) \cr
 & & -{5\over 2} \log(2\cos i),
\label{accretion-disk-Mv}
\end{eqnarray}
where $M_{\rm WD}$ is the WD mass, $\dot M_{\rm acc}$ the mass accretion
rate, and $i$ the inclination angle \citep[equation (A6) in][]{web87}.
Assuming that $M_{\rm WD}= 1.37 ~M_{\odot}$ and $i=80$\arcdeg
\citep{wou08}, we have $M_V =1.4$, 2.3, and 3.1 for the accretion rates
of $1 \times 10^{-6}$, $3 \times 10^{-7}$,
and $1 \times 10^{-7} ~M_{\odot}$~yr$^{-1}$, respectively. 

The apparent magnitude of the disk is calculated from 
\begin{equation}
m_V-M_V = A_V + 5 \log D_{10}, 
\label{distance-modulus}
\end{equation}
where $D_{10}$ is the distance divided by 10 pc.
With the absorption of $A_V = 1.6$ and the distance of 3 kpc, 
the apparent magnitude is estimated to be $m_V= 15.4$, 16.3 and 17.1,
for the above accretion rates,
respectively.  For the distance of 6.5 kpc,
we obtain brightnesses of $m_V = 17.1$, 18.0, and 18.8, 
respectively. All of these values are much fainter than 14.5 mag.
Therefore, it is very unlikely that an accretion disk mainly
contributes to the preoutburst luminosity.


\begin{figure}
\epsscale{1.15}
\plotone{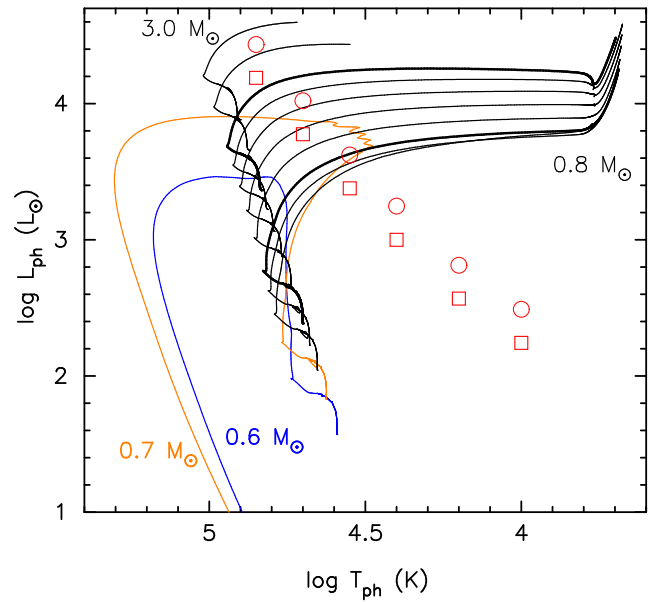}
\caption{
Evolutional tracks of helium stars with masses of $M_{\rm He}= 0.6$,
0.7, 0.8, 0.9, 1.0 (thick line), 1.2, 1.4, 1.6, 1.8, 2.0 (thick line),
2.5, and $3.0 ~M_\sun$ in the H-R diagram.
The 0.6 and $0.7~M_\sun$ stars evolve 
blueward and do not become red giants.  We stopped calculation when 
carbon ignites at the center for 2.5 and $3.0 ~M_\sun$ stars.
Open circles and squares denote stars with a 14.5 mag 
brightness at the distance of 6.5 kpc and 4.9 kpc, respectively, for
$A_V=1.6$. 
}
\label{hestarevolution}
\end{figure}

\subsection{Helium Star Companion}
Another possible source of the preoutburst brightness is
a helium star companion.  Figure \ref{hestarevolution} shows
evolutional tracks of helium stars with masses between 0.6 and
$3.0 ~M_\sun$ from the helium main-sequence to the red giant stage
in the H-R diagram.  We use OPAL opacities and an initial 
chemical composition of $X=0.0$, $Y=0.98$ and $Z=0.02$. 
The numerical method and input physics are the same as those
in \citet{sai95}. 

As shown in Figure \ref{hestarevolution},
low mass helium stars do not evolve to a red giant.
In our new calculation, the $0.8 ~M_{\odot}$ helium
star evolves to a red giant, but the 0.6 and $0.7 ~M_{\odot}$ helium
stars evolve toward blueward.  \citet{pac71} showed that  stars of
$M_{\rm He} \gtrsim 1.0 ~M_{\odot}$ evolve to a red giant while 0.5,
0.7, and $0.85 ~M_{\odot}$ do not.  Our calculations are essentially the 
same as those of \citet{pac71} and the difference is attributed
mainly to the difference between the adopted opacities. 

Figure \ref{hestarevolution} also shows locations of stars
whose apparent magnitudes are $m_V= 14.5$ for $A_V= 1.6$.
Here, the distance is assumed to be 4.9 (open squares) or 6.5 kpc
(open circles).  For example, a star of (temperature, luminosity) 
$= (\log T_{\rm ph} ~({\rm K})$,  $\log L_{\rm ph}/L_\sun) =$
(4.2, 2.81), (4.4, 3.25), and (4.6, 3.75) could be observed as a 14.5 mag
star for 6.5 kpc and $(\log T_{\rm ph} ~(K)$, 
$\log L_{\rm ph}/L_\sun) =$ (4.2, 2.57), (4.4, 3.0),
and (4.6, 3.51) for 4.9 kpc.  Therefore, the observed 14.5 mag is
consistent with the luminosities and temperatures of
slightly evolved helium stars of $M_{\rm He} \gtrsim 0.8 ~M_\sun$
if the companion is a blue star of $\log T_{\rm ph} \gtrsim 4.5$.

On the other hand, if the companion is redder than $\log T_{\rm ph}
\lesssim 4.4$, there is no corresponding evolution track
in the low luminosity region of the H-R diagram as shown in Figure
\ref{hestarevolution}.  In such a case the preoutburst luminosity 
cannot be attributed to a helium star companion. 
Kato (2001)\footnote{http://www.kusastro.kyoto-u.ac.jp/vsnet/Mail/alert5000/msg00493.html}
suggested that preoutburst color of V445 Pup was
much bluer than that of symbiotic stars (which have a red giant companion)
but rather close to that of cataclysmic variables (main-sequence or
little bit evolved companion). This argument is consistent with
our estimate of relatively high surface temperature
$\log T_{\rm ph} \gtrsim 4.5$ (see Fig. \ref{hestarevolution}).

\section{Discussions}

\subsection{Distance}
As explained in \S\S 3.1 and 3.2, we do not solve energy transfer outside
the photosphere and, thus, we cannot determine the proportionality
constant in equation (\ref{free-free-wind}).  Therefore, we cannot
determine the distance to V445 Pup directly from the comparison of 
theoretical absolute magnitude with the observational apparent magnitude. 
Instead, we can approximately estimate the distance by assuming
that the free-free flux is larger than the blackbody flux during
the observing period.  This assumption gives a lower limit to
the distance.  We expect that the distance estimated thus is 
close to the real value.  The 8th column of Table \ref{fitting_parameter}
lists the distance estimated for each model.  These distances are
consistent with the observational estimates of $3.5 \lesssim d \lesssim 6.5$ kpc 
\citep{iij08} and $\sim 4.9$ kpc \citep{wou08}.

\subsection{Helium Ignition Mass and Recurrence Period}


\begin{figure}
\epsscale{1.15}
\plotone{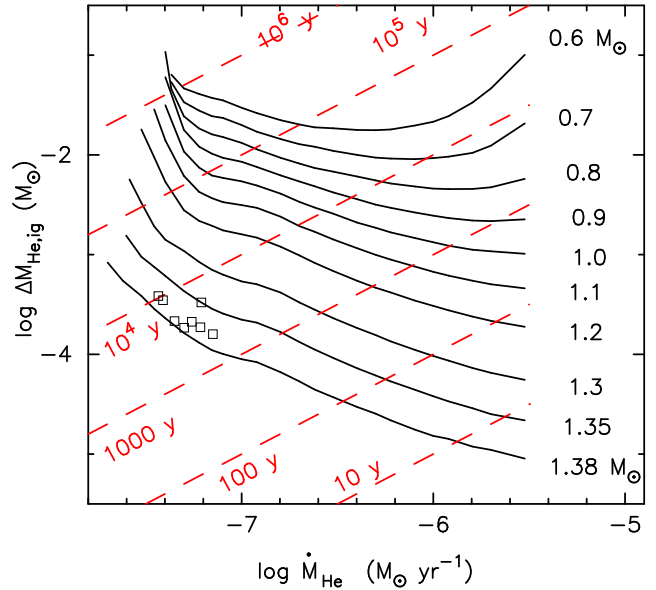}
\caption{
The helium ignition mass, $\Delta M_{\rm He, ig}$, of helium-accreting
WDs is plotted against the helium accretion rate, $\dot M_{\rm He}$.
The WD mass is attached to each curve. Straight dashed lines indicates
the recurrence period. The open square indicates 
the ignition mass of each model in Table \ref{fitting_parameter}.   
}
\label{dMenvMacc}
\end{figure}

We have calculated evolution of C+O white dwarfs accreting helium
at various rates until the ignition of a helium shell flash.
Chemical compositions assumed are $X_{\rm C}=0.48$ and 
$X_{\rm O}=0.50$ for the WD core, and $X=0.0$ and  $Y=0.98$ for
the accreted envelope.
The initial model adopted for a given core mass and accretion rate is
a steady-state model in which the heating due to the accretion is balanced
with the radiative energy flow. 
Adopting such an initial model is justified because the WD has accreted
matter from the companion for a long time and experienced many 
shell flashes.   

As the helium accretion proceeds, the temperature at the bottom of 
the envelope gradually increases. 
When the temperature becomes high enough the
triple-alpha reaction causes a shell flash.
The mass of the helium envelope at the ignition depends on the 
WD mass and the accretion rate as shown in Figure~\ref{dMenvMacc}.
The envelope mass required to ignite a shell flash tends 
to be smaller for a higher accretion rate and for a larger WD mass.
Helium flashes are weaker for higher accretion rates and lower WD masses.
In particular they are very weak for accretion rates higher than
$\sim 10^{-6} M_{\odot}$ yr$^{-1}$.

The envelope mass at the ignition depends also on the core temperature. 
If the core temperature
of the initial WD is lower than that of the steady state model
employed for our calculations, the envelope mass at the ignition
would be larger.    

The recurrence period,  $\Delta M_{\rm He,ig}/{\dot M}_{\rm He}$, is
also plotted in Figure~\ref{dMenvMacc} by dashed lines.
The recurrence period corresponding to the ignition mass of our 
fitted models are listed in Table \ref{fitting_parameter}.

\subsection{Mass Transfer from Helium Star Companion}

\begin{figure}
\epsscale{1.15}
\plotone{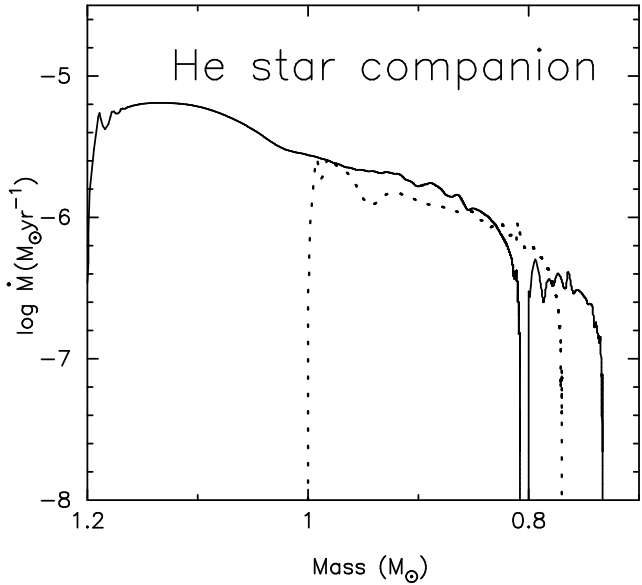}
\caption{
Mass loss rate of a helium star that fills its Roche lobe of the radius 
$1.5 ~R_\sun$. The initial stellar mass is 1.2, 1.0 (dotted line), and
$0.8 ~M_\sun$. Time goes on from left to right. See the text for more detail.
}
\label{Hestar.dmdt}
\end{figure}

We have estimated the mean accretion rate ${\dot M}_{\rm He}$ 
of the WD as in Table \ref{fitting_parameter}. 
The companion star feeds its mass to the WD via the Roche lobe overflow
or by winds depending on whether it fills the Roche lobe or not.
Here we examine if this accretion rate is comparable to the mass 
transfer rates of the Roche lobe overflow from a helium star
companion.  We followed stellar evolutions of helium stars from
the main sequence assuming that the star always fills its effective 
Roche lobe of which radius is simply assumed to be $1.5 ~R_\sun$. 
The resultant mass loss rates are shown in Figure \ref{Hestar.dmdt} 
for stars with initial masses of  1.2, 1.0 and $0.8 ~M_\sun$.
These results are hardly affected even if we change the Roche lobe
radius.  

The mass loss rate decreases with the companion mass almost
independently of the initial value. Except for 
the short initial and final stages, these rates are
as large as $\dot M_{\rm He} \sim 10^{-6} ~M_\sun$~yr$^{-1}$ 
and much larger than the mass accretion rate estimated from our 
fitted models in Table \ref{fitting_parameter}.
Since almost all of the mass lost by the companion accretes
onto the WD in the case of Roche lobe overflow,
the accretion may result in very weak shell flashes 
for such high rates as $10^{-6} M_\sun$~yr$^{-1}$.
A strong shell flash like V445 Pup is realized only when 
the final stages of mass-transfer in which the mass transfer rate 
quickly drops with time. This rare case may correspond to a 
final stage of binary evolution after the WD had grown up in mass to 
near the Chandrasekhar limit from a less massive one and the companion
had lost a large amount of mass via Roche lobe overflow. 
Such information will be useful for modeling a new path of 
binary evolution scenario that will lead to a Type Ia supernova.


\subsection{Comparison with Other Observation}
The mass of the dust shell can be estimated from infrared continuum flux 
with the assumption that the emission originated from warm dust.
With infrared 10 $\mu$m spectrum, \citet{lyn04b} estimated the dust 
mass to be $2 \times 10^{-6} M_{\odot}$ for a distance of 3 kpc. 
This value would be increased if we adopt a larger distance
instead of 3 kpc, but still consistent with our estimated 
ejecta mass $\Delta M_{\rm ej}\sim (0.7-1.8) \times 10^{-4} 
M_{\odot}$ in Table \ref{fitting_parameter}, 
because the dust mass is a small part of the ejected mass. 

\citet{lyn04a} reported the absence of He II or coronal lines on 
their near IR spectra and suggested that the ionizing source was
not hot enough on 2004 January 14 and 16 (1146 days after 
the optical maximum).   \citet{lyn05} also reported,
from their near IR observation on 2005 November 16
that the object had faded and the thermal dust emission 
had virtually disappeared. They suggested that the dust had 
cooled significantly until that date (1818 days).  This suggestion
may constrain the WD mass of V445 Pup because
Figure \ref{lightcurve} shows the outburst lasts more than
1800 days for less massive WDs ($ \lesssim 1.33 ~M_\sun$). 
If the WD had cooled down until the above date we may exclude
less massive WD models ($ \lesssim 1.33 ~M_\sun$) because  
these WDs evolve slowly and their hot surfaces emit high energy photons 
at least until 1800 days after the optical peak.

We have shown that the WD mass of V445 Pup is increasing with time.
When the WD continues to grow up to the Chandrasekhar limit,
central carbon ignition triggers a Type Ia supernova explosion 
if the WD consists of carbon and oxygen. We regards the star as 
a CO WD instead of an O-Ne-Mg WD because no indication of neon 
were observed in the nebula phase spectrum \citep{wou05}. Therefore, 
we consider that V445 Pup is a strong candidate of Type Ia supernova.

When a binary consisting of a massive WD and a helium star
like V445 Pup becomes a Type Ia supernova, its spectrum may more or less 
show a sign of helium.  The search for such helium associated with
a Type Ia supernova has been reported for a dozen Type Ia supernovae 
\citep{mar03, mat05}, but all of them are negative detection.
It may be difficult to find such a system because this type of
Type Ia supernovae are very rare.

The binary seems to be still deeply obscured by the optically thick
dust shell even several years after the outburst. This blackout period
is much longer than that of the classical novae such as OS And ($\sim 20$
days), V705 Cas ($\sim 100$ days), and DQ Her ($\sim 100$ days). 
As the ejected mass estimated in Table \ref{fitting_parameter} is
not much different from those of classical novae, the difference of 
the blackout period may be attributed to a large amount of dust in
the extremely carbon-rich ejecta of a helium nova.
Moreover, observed low ejection velocities of $\sim 500$ km~s$^{-1}$
\citep{iij08} is not much larger than the escape velocity of
the binary with a relatively massive total mass [e.g., $1.35 ~M_\sun
+ (1 - 2)  ~M_\sun$].  Both of these effects lengthen the dust
blackout period.  When the dust obscuration will be cleared
in the future the period of blackout provides useful information on 
the dust shell.

\section{Conclusions}
Our main results are summarized as follows:

1. We  have reproduced $V$ and $I_{\rm c}$ light curves
using free-free emission dominated light curves calculated on the 
basis of the optically thick wind theory.

2. From the light curve analysis we have estimated the WD mass
to be as massive as $M_{\rm WD} \gtrsim 1.35~M_\sun$.

3. Our light curve models are now consistent with a longer distance
of $3.5 \lesssim d \lesssim 6.5$ kpc \citep{iij08} and 4.9 kpc \citep{wou08}. 

4. We have estimated the ejecta mass as the mass lost by optically
thick winds, i.e., $\Delta M_{\rm wind} \sim 10^{-4} M_\odot$.
This amounts to about a half of the accreted helium matter so that
the accumulation efficiency reaches $\sim 50$\%. 

5. The white dwarf is already very close to the Chandrasekhar mass,
i.e., $M_{\rm WD} \gtrsim 1.35~M_\sun$, and the WD mass had increased
after the helium nova.  Therefore, V445 Pup is a strong candidate
of Type Ia supernova progenitors.

6. We emphasize importance of observations after the dense
dust shell will disappear, especially observations of the color and
magnitude, orbital period, and inclination angle of the orbit.
These are important to specify companion nature.



\acknowledgments 
M.K. and I.H. are grateful to people at the Astronomical Observatory of 
Padova and at the Department of Astronomy of the University of Padova, Italy,
for their warm hospitality.  Especially we thank Takashi Iijima for
fruitful discussion on V445 Pup, which stimulated us to start this work. 
The authors thank Masaomi Tanaka for information on the helium detection 
in Type Ia supernovae. We also thank the anonymous referee for 
useful comments to improve the manuscript.  Thanks are also due to
the American Association of Variable Star Observers (AAVSO)
for the photometric data of V445 Pup and Taichi Kato for introducing us
the discussion in VSNET.  This research has been supported
in part by the Grants-in-Aid for Scientific Research (16540211, 20540227)
from the Japan Society for the Promotion of Science.

{}

\end{document}